\documentclass[letter,5p,twocolumn,amssymb,amsmath]{elsarticle}
\usepackage[breaklinks]{hyperref}

\usepackage{url}

\usepackage{breakurl}

\usepackage{graphicx}
\usepackage{epstopdf}
\usepackage{braket}
\usepackage{bm}

\hypersetup{colorlinks,allcolors=black}

\journal{Journal of \LaTeX\ Templates}

\begin{document}

\begin{frontmatter}

\title{Novel Penning-trap techniques reveal isomeric states\\ in $^{128}$In and $^{130}$In for the first time}


\author[jyu]{D.A.~Nesterenko}
\author[jyu]{A.~Kankainen}
\author[jyu]{J.~Kostensalo}
\author[brighton]{C.R.~Nobs}
\author[brighton]{A.M.~Bruce}
\author[jyu]{O.~Beliuskina}
\author[jyu]{L.~Canete}
\author[jyu]{T.~Eronen}
\author[brighton]{E.R.~Gamba}
\author[jyu]{S.~Geldhof}
\author[jyu]{R.~de~Groote}
\author[jyu]{A.~Jokinen}
\author[warsaw]{J.~Kurpeta}
\author[jyu]{I.D.~Moore}
\author[surrey]{L.~Morrison}
\author[surrey]{Zs.~Podoly\'ak}
\author[jyu]{I.~Pohjalainen}
\author[jyu]{S.~Rinta-Antila}
\author[jyu]{A.~de~Roubin}
\author[surrey]{M.~Rudigier}
\author[jyu]{J.~Suhonen}
\author[jyu]{M.~Vil\'en}
\author[jyu]{V.~Virtanen}
\author[jyu]{J.~\"Ayst\"o}

\address[jyu]{University of Jyvaskyla, P.O. Box 35, FI-40014 University of Jyvaskyla, Finland}
\address[brighton]{School of Computing, Engineering and Mathematics, University of Brighton, Brighton BN2 4GJ, United Kingdom}
\address[surrey]{Department of Physics, University of Surrey, Guildford GU2 7XH, United Kingdom}
\address[warsaw]{Faculty of Physics, University of Warsaw, ulica Pasteura 5, PL-02-093 Warsaw, Poland}

\begin{abstract}
Isomeric states in $^{128}$In and $^{130}$In have been studied with the JYFLTRAP Penning trap at the IGISOL facility. By employing novel ion manipulation techniques, different states were separated and masses of six beta-decaying states were measured. JYFLTRAP was also used to select the ions of interest for identification at a post-trap decay spectroscopy station. A new beta-decaying high-spin isomer feeding the $15^-$ isomer in $^{128}$Sn has been discovered in $^{128}$In at $1797.6(20)$~keV. Shell-model calculations employing a CD-Bonn potential re-normalized with the perturbative G-matrix approach suggest this new isomer to be a $16^+$ spin-trap isomer. In $^{130}$In, the lowest-lying $(10^-)$ isomeric state at $58.6(82)$~keV was resolved for the first time using the phase-imaging ion cyclotron resonance technique. The energy difference between the $10^-$ and $1^-$ states in $^{130}$In, stemming from parallel/antiparallel coupling of $(\pi 0g_{9/2}^{-1})\otimes(\nu 0h_{11/2}^{-1})$, has been found to be around 200 keV lower than predicted by the shell model. Precise information on the energies of the excited states determined in this work is crucial for producing new improved effective interactions for the nuclear shell model description of nuclei near $^{132}$Sn.
\end{abstract}


\end{frontmatter}

Neutron-rich indium isotopes provide essential data to test the nuclear shell model \cite{Mayer1949,Haxel1949} and to further develop nucleon-nucleon interactions and related potentials \cite{Caurier2005,Corragio2009}. This is important for example to obtain better predictions for the astrophysical rapid neutron capture process \cite{Horowitz2019} traversing through the $N=82$ isotones and forming its second abundance peak at $A\approx130$. Recently, many decay spectroscopy experiments have been performed in the $A\approx130$ region \cite{Taprogge2014,Taprogge2015,Jungclaus2016a,Jungclaus2016b,Lorenz2019,Chen2019,Phong2019}, and given evidence e.g. for a reduction of the $Z=40$ proton subshell gap when approaching $N=82$ \cite{Taprogge2014,Chen2019}. Despite these advances, excitation energies for many long-living beta-decaying isomeric states have remained unknown although they can provide crucial information on the nucleon-nucleon interactions close to $^{132}$Sn. 

Isomers have a different spin, shape, or structure compared to the lower-lying states in the nucleus (see e.g.~ \cite{Walker1999}), hindering their de-excitation and prolonging the lifetimes. High-spin isomers in odd-odd nuclei, such as $^{128}$In studied in this work, cannot be populated via the ground-state beta decay of their even-even parent nucleus. The fission yields of high-spin isomers can also be lower than for the ground states, making it possible to miss related beta decays or even the existence of such isomers. For example the isomeric yield fraction of the ($21/2^-$) isomer in $^{127}$In has been measured to be less than 30~\% in proton-induced fission on uranium \cite{Rakopoulos2019}. In this work, we employ novel Penning-trap techniques to study the  beta-decay of isomeric states in background-free conditions giving key information on the excited states they populate in the daughter nucleus. 


The pioneering work on the isomeric states of neutron-rich indium isotopes done by Fogelberg \emph{et al.} \cite{Fogelberg1974,Fogelberg1979} focused on the even-$A$ isotopes $^{120-128}\rm In$. These studies have recently been extended, at RIKEN, to $^{130}\rm In$ \cite{Jungclaus2016b} and $^{132}\rm In$ \cite{Jungclaus2016a} studied via the beta decay of $^{130,132}$Cd using the EURICA detector setup and to $^{134}\rm In$ \cite{Phong2019} populated in the in-flight fission of $^{238}$U. Penning-trap mass spectrometry offers a way to determine the excitation energies of long-living isomeric states as has been done for the odd-$A$ $^{129}$In and $^{131}$In isotopes at JYFLTRAP \cite{Kankainen2013}. In addition to low-spin isomers, high-spin isomers with spin parities $23/2^-$ and $21/2^-$ have been observed in the $^{129,131}$In isotopes \cite{Gausemel2004, Fogelberg2004}. Recently, several indium isotopes were measured with the TITAN Penning trap \cite{Babcock2018} but some isomeric states were not fully resolved.



For even-$A$ nuclei the high-spin states of $^{122,124,126}$In have been recently studied using the nuclear shell model \cite{Rejmund2016}. A good agreement between the shell-model calculations and the experiment was found using the effective interaction jj45pna. On the other hand, restrictions to the model space had to be used for the lighter isotopes which needed to be compensated by re-adjusting the effective charges of the nucleons. As computational power increases, calculations in the full relevant model space become possible. Information on the spins, parities, energies, and reduced transition probabilities are vital for fitting new effective interactions for these previously computationally problematic model spaces. The present paper is a step towards understanding the properties of nuclei south-west of $^{132}\rm Sn$. 

In this work, we have studied long-living beta-decaying states in $^{128}$In and $^{130}$In by applying novel ion-trapping methods to measure their masses and decay properties. The neutron-rich indium isotopes were produced with a 30-MeV proton beam impinging into a uranium target at the Ion Guide Isotope Separator On-Line (IGISOL) facility \cite{Moore2013}. The fission fragments were thermalized in helium gas, extracted from the IGISOL gas cell and guided towards the high-vacuum region of the mass separator using a sextupole ion guide \cite{Karvonen2008}. Most of the fragments end up as singly-charged ions, which were accelerated to 30 keV, and mass-separated with a dipole magnet. The continuous $A/q$ beam was cooled and bunched employing the radiofrequency quadrupole cooler and buncher (RFQ) \cite{Nieminen2001} before injecting into the JYFLTRAP double Penning trap mass spectrometer \cite{Eronen2012}. A dedicated post-trap spectroscopy setup was prepared after JYFLTRAP to identify the states whose masses had been studied. The isomerically purified ion bunches from JYFLTRAP were implanted into a movable mylar tape surrounded by a scintillator detector, two 70~\% coaxial and a broad-energy range Ge detector. 

\begin{figure}[!htb]
\centering
\includegraphics[width=0.4\textwidth,clip]{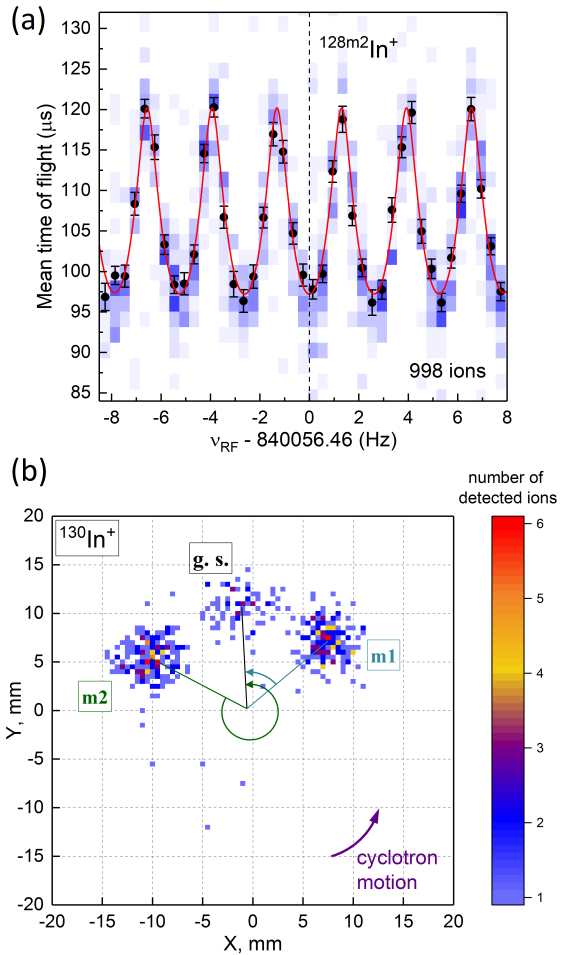}
\caption{(Color online) Typical measurements performed at JYFLTRAP. (a) TOF-ICR spectrum for $^{128}$In$^{m2+}$ with 25-350-25 ms (On-Off-On) Ramsey excitation pattern. The isomeric state was selected by using a Ramsey dipolar cleaning method before the actual measurement. The black points with error bars represent the mean time-of-flight for each scanned frequency. The solid red line is a fit of the theoretical
curve to the data points \cite{Kretzschmar2007}. The blue shading around the data points indicates the number of ions in each time-of-flight bin. (b) Projection of the cyclotron motion of $^{130}$In$^+$ ions onto the position-sensitive detector obtained with the PI-ICR technique using a 320~ms phase accumulation time.}
\label{fig:piicr} 
\end{figure}

At JYFLTRAP, the ions were first cooled and purified using the buffer-gas cooling technique \cite{Savard1991} in the first trap. It allowed to clean the ions from isobaric contaminants. To resolve the isomeric states from each other and from the ground state an additional purification step employing a Ramsey dipolar cleaning \cite{Eronen2008} pattern with two 5-ms excitation fringes, separated by either 40 ms ($^{128}$In$^{m2}$ and $^{130}$In) or 90 ms ($^{128}$In and $^{128}$In$^{m1}$) waiting time in between, was applied in the second trap. This was further followed by a cooling period in the first trap before the actual mass measurements using the time-of-flight ion cyclotron resonance (TOF-ICR) \cite{Graff1980,Konig1995} technique in the second trap to determine the ion's cyclotron frequency $\nu_c=qB/(2\pi m)$, where $q$ and $m$ are the charge and the mass of the ion and $B$ is the magnetic field strength. The measurements were performed using time-separated oscillatory fields \cite{Kretzschmar2007,George2007} with 25 ms (On) - 350 ms (Off) - 25 ms (On) pattern for $^{128}$In (see Fig.~\ref{fig:piicr}.(a)) and 25 ms (On) - 150 ms (Off) - 25 ms (On) pattern for $^{130}$In. The magnetic field strength was determined using $^{128}$Te (mass excess $\Delta=-88993.7(9)$~keV \cite{AME16}) and $^{130}$Te ($\Delta=-87352.949(11)$~keV \cite{AME16}) as references for $^{128}$In and $^{130}$In, respectively. The use of isobaric references had the benefit that possible systematic uncertainties due to imperfections in the trap cancel out \cite{Roux2013}. Time-dependent fluctuations in the magnetic field strength \cite{Canete2016} were also taken into account in the analysis. Count-rate class analysis \cite{Kellerbauer2003} was performed to account for ion-ion interactions in the trap. For the final result, a weighted mean and its inner and outer errors \cite{Birge1932} were calculated, and the larger of the errors was adopted. The results from the TOF-ICR measurements are summarized in Table~\ref{tab:results} and noted with $^a$. 

For $^{128}$In, it is interesting to compare the results obtained in this work with clean samples of $^{128}$In$^+$ states at JYFLTRAP, to the combined ground state and isomer measurements performed with $^{128}$In$^{13+}$ ions using the TITAN Penning trap at TRIUMF \cite{Babcock2018}. Whereas the masses for $^{128}$In$^{m1}$ agree well between the two measurements, the ground-state mass determined from the two-state fit on $^{128}$In$^{13+}$ \cite{Babcock2018} is 20(10)~keV higher than the seven times more precise JYFLTRAP value. As a result, the excitation energy obtained at TITAN is 23(13)~keV lower than the JYFLTRAP value (see Table~\ref{tab:results}).

For $^{130}$In, the TOF-ICR measurement was not able to resolve the $1^{(-)}$ ground state from the $(10^-)$ isomeric state lying at $50(50)$~keV \cite{Singh2001}. The TOF-ICR resonances collected with 400 ms and 600 ms excitation times showed a similar production ratio between the $(5^+)$ isomer and the lower-mass state. This suggests that the lower-mass state was the $(10^-)$ level which has a similar half-life to the $(5^+)$ isomeric state (see Table~\ref{tab:results}) whereas the $1^{(-)}$ ground state has a much shorter half-life of 290(20)~ms \cite{Singh2001}. This is consistent with the non-observation of the most prominent gamma lines from the beta decay of the $1^{(-)}$ state in the collected beta-gated gamma-ray spectra for the studied lower-mass state. We conclude that the $(10^-)$ and $(5^+)$ isomers in $^{130}$In were measured with the TOF-ICR technique, however, the result for the $(10^-)$ state might still contain a small contribution from the weakly produced ground state.

\begin{table*}
\centering
\caption{\label{tab:results} Isomeric states in $^{128}$In and $^{130}$In studied in this work together with their spins, parities $J^{\pi}$, and half-lives $T_{1/2}$ from literature \cite{Singh2001,Elekes2015}. The frequency ratios $r=\nu_{c,ref}/\nu_{c}$ determined using the TOF-ICR ($^a$) and PI-ICR ($^b$) techniques in this work, corresponding mass-excess values $\Delta$ and excitation energies $E_{x}$ are tabulated and compared to the literature values from \cite{Babcock2018,Singh2001}. The reference nuclides have been listed for each measurement. The TOF-ICR measurement of $^{130}$In$^{m1}$ (marked with $^c$) was done as an admixture with the ground state, and hence the PI-ICR value is recommended.}
\begin{tabular}{lllllllll}
\hline\noalign{\smallskip}
Nuclide & $J^{\pi}$ & $T_{1/2}$ & Ref. & $r=\nu_{c,ref}/\nu_{c}$ & $\Delta$  & $\Delta_{lit.}$  & $E_x$  & $E_{x,lit.}$ \\
 &  &  (s)  &  &  & (keV) & (keV) & (keV) & (keV)\\
\hline\noalign{\smallskip}
$^{128}$In 				& $(3)^+$  & 0.84(6) 		& $^{128}$Te   			& $1.000124253(9)^a$ 		& $-74190.0(14)$ 	& $-74170.5(97)$ & 0.0						&	\\
$^{128}$In$^{m1}$ & $(8^-)$  & 0.72(10) 	& $^{128}$Te 				& $1.0001266459(16)^a$ 	& $-73904.9(21)$ 	& $-73908.8(91)$ & $285.1(25)$ 	& $262(13)$ \\
$^{128}$In$^{m2}$ & $(16^+)$ & $\geq 0.3$ & $^{128}$Te 				& $1.000139341(10)^a$		& $-72392.4(15)$	& - 						 & $1797.6(20)$ & -\\
\hline
$^{130}$In 				& $1^{(-)}$ & 0.29(2) 	& $^{130}$In$^{m2}$ & $0.999996815(50)^b$		& $-69909.2(75)$ & $-69862(20)$  & 0.0							& \\
\hline
$^{130}$In$^{m1}$ & $(10^-)$  & 0.54(1)		& $^{130}$Te 				& $1.000144705(53)^{a,c}$	 		& $-69842.7(64)$ & 		& $66.5(99)$ & \\
									& 					&						& $^{130}$In$^{m2}$ & $0.999997292(46)^b$  	& $-69850.6(71)$ & $-69830(40)$		& $58.6(82)$ & $50(50)$  \\
\hline
$^{130}$In$^{m2}$ & $(5^+)$ & 0.54(1)   	& $^{130}$Te			  & $1.000147325(41)^a$ 	& $-69525.6(49)$ & 							& $384(9)$ 				& \\
									& 				& 						& $^{133}$Cs				& $0.977577292(36)^b$ & $-69522.8(45)$ & 							& $386.3(60)$ & \\
									& 				& 						& 									& Weighted mean:									& $-69524.1(33)$ & $-69503(28)$ & $385.5(50)$ & $359(34)$ \\
\noalign{\smallskip}\hline
\end{tabular}
\end{table*} 

To resolve all three states in $^{130}$In, a novel phase-imaging ion cyclotron resonance (PI-ICR) technique \cite{Eliseev2013,Eliseev2014,Nesterenko2018} was employed at JYFLTRAP. The cyclotron frequency was determined based on the phase difference after a phase accumulation time $t_{acc}$. With the PI-ICR technique, all three states were resolved (see Fig.~\ref{fig:piicr}.(b)). The $^{130}$In$^{m2}$ isomer was measured using $^{133}$Cs ($\Delta=-88070.931(8)$~keV \cite{AME16}) as a reference ($t_{acc}=250$~ms), and the other two states were measured against $^{130}$In$^{m2}$ with $t_{acc}$=320~ms. The data analysis followed otherwise the same procedures as described for TOF-ICR measurements. The PI-ICR frequency ratio results are highlighted with $^b$ in Table~\ref{tab:results}.

The shorter-living ground state of $^{130}$In was the least populated in the PI-ICR spectra and supports the conclusion that the lower-mass state in the TOF-ICR measurements was predominantly the $(10^-)$ state. The mass-excess values determined from the TOF-ICR and PI-ICR measurements of $^{130}$In$^{m1}$ and $^{130}$In$^{m2}$ agree with each other (see Table~\ref{tab:results}). The recent measurement at the TITAN Penning trap \cite{Babcock2018} reports mass-excess values of $-69862(20)$~keV for the $1^{(-)}$ ground state and $-69503(28)$~keV for the $(5^+)$ isomer in $^{130}$In. The value reported for the $1^{(-)}$ ground state from TITAN \cite{Babcock2018} is well above the value from this work and it actually fits better with the $(10^-)$ state measured in this work. The ground-state mass determined in this work, $-69909.2(75)$~keV, agrees well with AME16 ($-69880(40)$~keV \cite{AME16}).

\begin{figure}[htb]
\centering
\includegraphics[width=0.49\textwidth,clip]{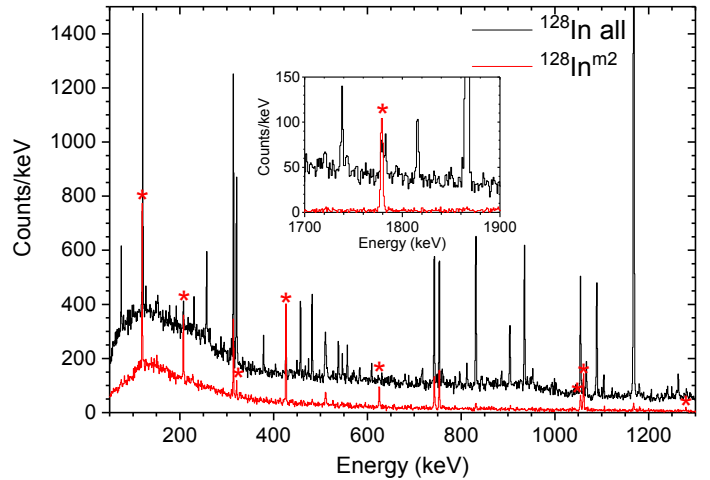}
\caption{(Color online) Beta-gated gamma-ray spectrum obtained with the trap purification set to $^{128}$In (including all three states, in black) and to the new high spin isomer $^{128}$In$^{m2}$ at 1.8 MeV (in red). The peaks tabulated in Table~\ref{tab:gammas} are shown with an asterisk (*). The strong 1779-keV transition is shown in the inset.
\label{fig:128In} }
\end{figure}

A new isomeric state in $^{128}$In at an excitation energy of $1797.6(20)$~keV was discovered in this work. The yield for this new isomer was similar to the first isomeric state in $^{128}$In, and both had trap cycles of around 0.75~s. Therefore, it is estimated that the new isomeric state $^{128}$In$^{m2}$ has to have a half-life longer than 0.3~s. Since the state was previously unknown, a pure beam of $^{128}$In$^{m2}$ was prepared with the trap, and implanted on a tape which was moved after every 1000~seconds ($\approx 17$~mins). For comparison, a spectrum employing only first-trap purification not sufficient to resolve the three states, was also collected. Figure~\ref{fig:128In} shows the two beta-gated gamma-ray spectra obtained with these settings. 

\begin{figure}[htb]
\centering
\includegraphics[width=0.8\linewidth,clip]{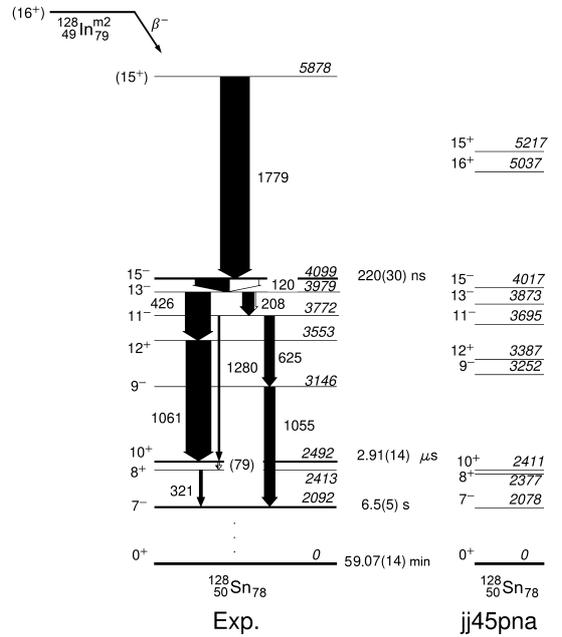}
\caption{Levels in $^{128}$Sn fed by the beta decay of $^{128}$In$^{m2}$ ($J^{\pi}=(16^+)$, $Q_\beta=10970(18)$~keV) studied in this work (left). For comparison, level scheme based on shell-model calculations using the effective interaction jj45pna \cite{jj45pna} is given (right). Only transitions down to the $7^-$ isomer are shown. The 79 keV transition marked in parentheses was not detected.
\label{fig:sn128levels}}
\end{figure}

Most of the observed beta-delayed gamma transitions from the new $^{128}$In$^{m2}$ and their intensities match with the transitions observed from the $15^-$ isomer with $T_{1/2}$ = 220(30)~ns in $^{128}$Sn in Refs.~\cite{Pietri2011,Iskra2014}. Therefore, the new isomeric state in $^{128}$In has to populate the $(15^-)$ isomer in $^{128}$Sn either directly or indirectly. We also observe two gamma transitions (1280 keV and 1779 keV) not observed in \cite{Pietri2011,Iskra2014}. Of these, the strong 1779 keV transition has an intensity similar to the $15^- \rightarrow 13^-$ transition, suggesting it feeds the $15^-$ isomer. The 1779 keV transition has been already observed in \cite{Fogelberg1979} where it was not assigned because it was not coincident with other gamma transitions within the used time window of 2-20~ns. Therefore, coincidences between the transitions above and below the $15^-$ state with a half-life of 220(30)~ns could not have been observed in Ref.~\cite{Fogelberg1979}. The resulting level scheme for $^{128}$Sn is shown left in Fig.~\ref{fig:sn128levels}.
 

\begin{table}
\centering
\caption{\label{tab:gammas} Observed $\gamma$ transitions following the beta decay of $^{128}$In$^{m2}$ and their relative intensities. The conversion coefficients $\alpha$ from the BrIcc calculator \cite{Kibedi2008} were used to obtain the total intensity $I_{tot}=(1+\alpha)I_{\gamma}$. This mainly concerns the 119.5-keV ($\alpha=0.848(12)$) and 207.5-keV ($\alpha=0.1226(18)$) $E2$ transitions since for the others, $\alpha< 0.01$. Due to the used coincidence gate of 1.5~$\mu$s, the $\gamma$ transitions following the $10^+$ ($T_{1/2}=3$~$\mu$s) and $7^-$ ($T_{1/2}=6.5(5)$~s) states were strongly suppressed (marked with $^*$). For comparison, intensities obtained within 1.5 $\mu$s after the $^{128}$Sn implantation in Ref.~\cite{Pietri2011} are given, renormalized to the 119-keV transition.}
\begin{tabular}{llll}
\hline\noalign{\smallskip}
$E_\gamma$ & $I_{tot}$ (\%) & $I_{tot}$ (\%) \cite{Pietri2011} & $J^\pi_i \rightarrow J^\pi_f$ \\
\hline\noalign{\smallskip}
119.5(3)		&	100(8) & 100(14)&	$15^- \rightarrow 13^-$	\\
207.5(3)		&	34(4)	& 29(7) &	$13^- \rightarrow 11^-$	\\
321.1(4)$^*$	  &	9(2)	&	& $8^+ \rightarrow 7^-$	\\
426.0(3)		&	74(8)	& 53(9) & $13^- \rightarrow 12^+$	\\
625.4(3)		&	29(4)	& 25(6)	& $11^- \rightarrow 9^-$	\\
831.3(2)$^*$ & 9(2) & & $4^+ \rightarrow 2^+$ \\
1054.8(3)	  &	31(5)	& 17(6)	& $9^- \rightarrow 7^-$	\\
1061.1(3)	  &	73(8)	& 62(11)	& $12^+ \rightarrow 10^+$	\\
1168.3(2)$^*$ & 13(3) & & $2^+ \rightarrow 0^+$ \\
1279.8(5)		&	5(2)	& & ($11^- \rightarrow 10^+$) \\
1779.1(3)	  &	85(8) &  & ($(15^+) \rightarrow 15^-$)\\
\noalign{\smallskip}\hline
\end{tabular}
\end{table} 


\begin{figure}[htb]
\centering
\includegraphics[width=0.49\textwidth,clip]{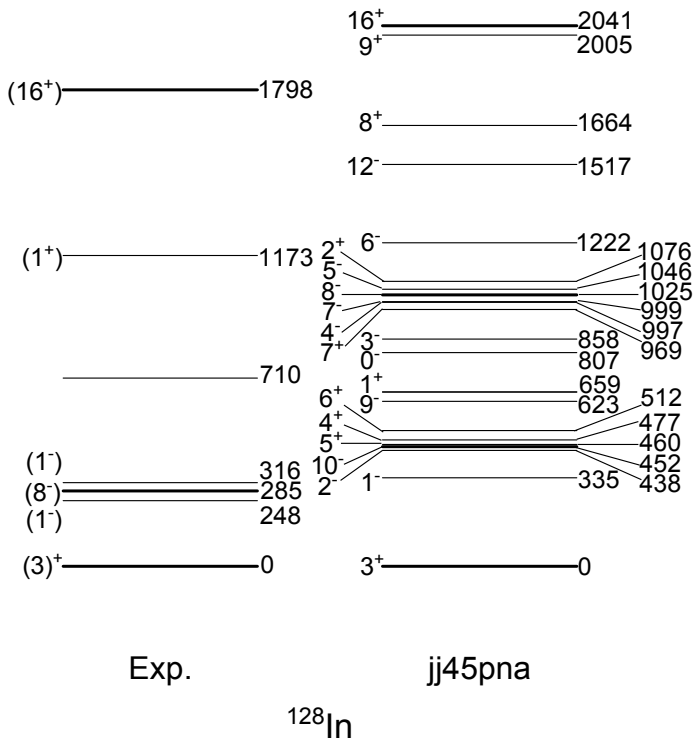}
\caption{Experimental level scheme of $^{128}$In based on this work and \cite{nndc} (left). The lowest excited states for each spin-parity were calculated with the shell model using the effective interaction jj45pna \cite{jj45pna} (right). The calculated excitation spectrum contains also many other states with the same spin-parities but they are too numerous to be presented in this figure. In fact, the $16^+$-state is the 71st state in $^{128}\rm In$. }
\label{fig:128inshell} 
\end{figure}

\begin{figure*}[htb]
\centering
\resizebox{1\textwidth}{!}{%
  \includegraphics{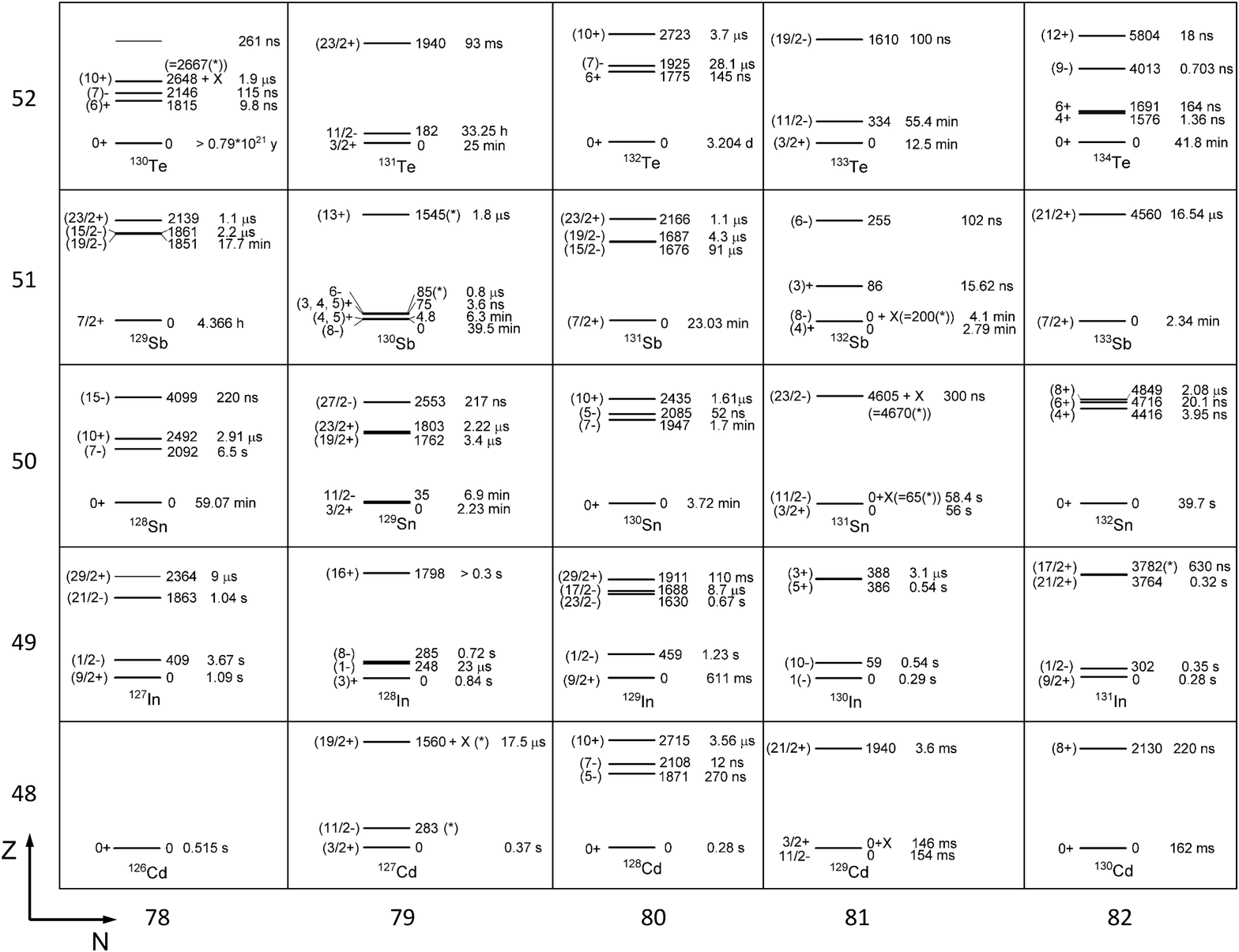}
}
\caption{Systematics of the isomeric states in the investigated region. Data are taken from ENSDF and XUNDL (marked with (*)) \cite{nndc} and include the excitation energies of the isomeric states in $^{128}$In and $^{130}$In from this work.}
\label{fig:systematics} 
\end{figure*}

\begin{figure}[htb]
\centering
\includegraphics[width=0.4\textwidth,clip]{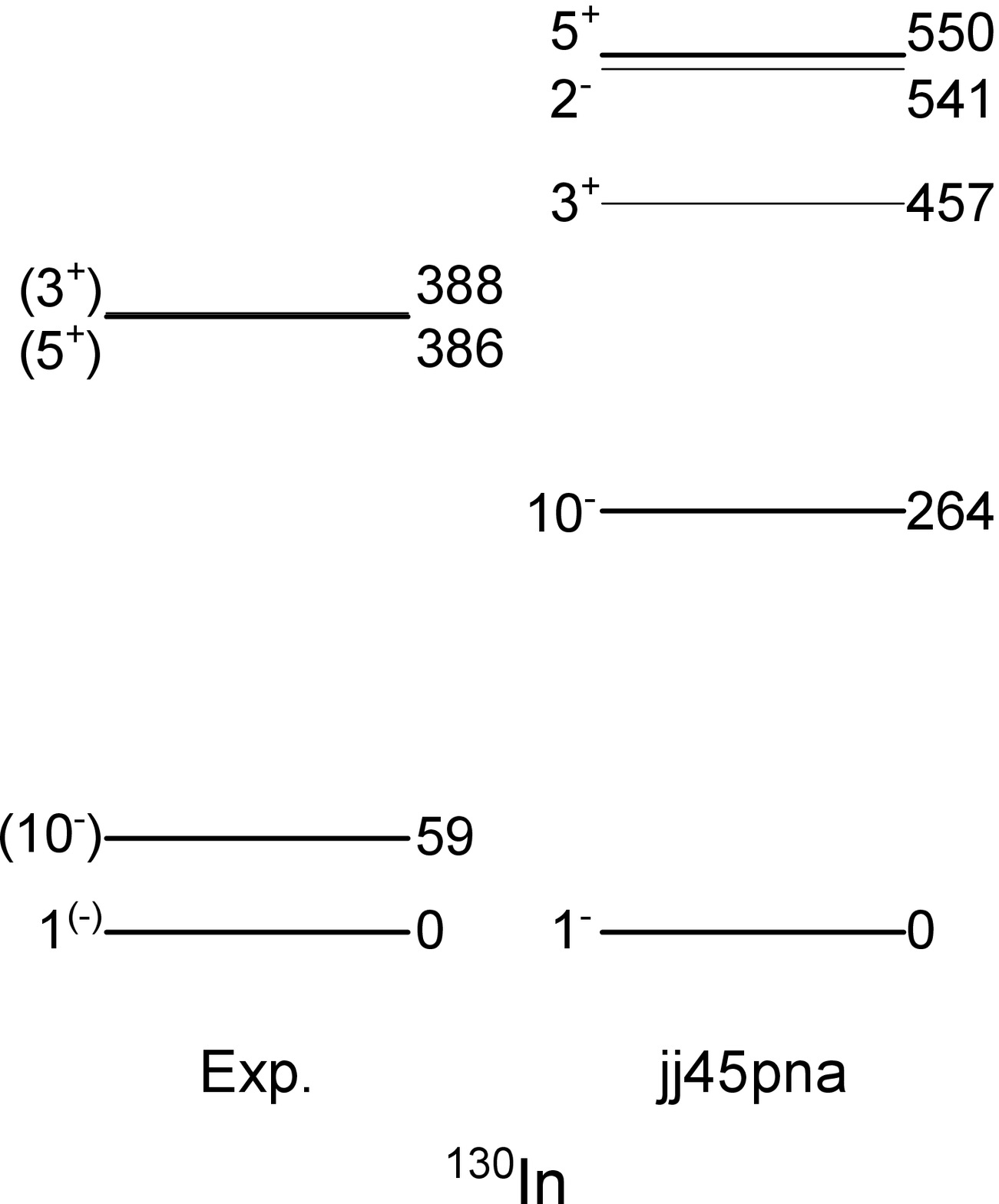}
\caption{The experimental level scheme of $^{130}$In based on this work, together with the $3^+$ state from \cite{nndc}, compared with the shell-model calculations using the effective interaction jj45pna \cite{jj45pna}. The calculated level scheme contains only the lowest excited states for each spin-parity. There are also many other states with the same spin-parities but they are too numerous to be presented in this figure.}
\label{fig:130inshell} 
\end{figure}

To further investigate the studied states and the new high-spin isomer in $^{128}$In, shell-model calculations were performed in a valence space consisting of the proton orbitals $1p_{3/2}$, $0f_{5/2}$, $1p_{1/2}$, and $0g_{9/2}$, and the neutron orbitals $0g_{7/2}$, $1d_{5/2}$, $1d_{3/2}$, $2s_{1/2}$, and $0h_{11/2}$ using the shell model code NuShellX@MSU \cite{nushellx} with the effective interaction jj45pna \cite{jj45pna}. The interaction jj45pna is a CD-Bonn potential re-normalized with the perturbative G-matrix approach. Interestingly, the calculations predict that the first isomeric state in $^{128}$In would be $10^-$ (see Fig.~\ref{fig:128inshell}), similar to $^{130}$In, but in disagreement with experiments suggesting it is $(8^-)$ \cite{nndc} based on the observed feeding to a $(7^-)$ state in the daughter. However, the feeding might also be explained by the missed transitions from higher-lying levels. The excitation energy determined here for the first isomeric state in $^{128}$In, $285.1(25)$ keV, is significantly higher than recently obtained at TITAN (262(13)~keV \cite{Babcock2018}) but still lower than the theoretical prediction (452 keV).

According to the shell-model calculations, the new high-spin isomer in $^{128}$In is $16^+$ since no other spin-trap states are located at around 2 MeV (see Fig.~\ref{fig:128inshell}). The $16^+$ state consists 92~\% of the configuration $(\pi 0g_{9/2})^{-1}\otimes(\nu 1d_{3/2}^{-1}0h_{11/2}^{-2})$. The $16^+$ assignment is further supported by the systematics of high-spin isomers in the $N=79$ isotones $^{129}$Sn, $^{130}$Sb and $^{131}$Te (see Fig.~\ref{fig:systematics}). They all have high-spin isomers with similar leading neutron configurations $\nu 1d_{3/2}^{-1}0h_{11/2}^{-2}$, located at 1803 keV ($(23/2^+)$ in $^{129}$Sn \cite{Genevey2002,Gausemel2004}), 1545 keV ($(13^+)$ in $^{130}$Sb \cite{Genevey2002}) and 1940 keV ($(23/2^+)$ in $^{131}$Te \cite{Zhang1998}). In addition, similar high-spin isomers are also found in neighbouring indium isotopes, such as the $(21/2^-)$ state in $^{127}$In and $(23/2^-)$ in $^{129}$In \cite{Gausemel2004}.

Allowed beta decay from the new $(16^+)$ isomer would populate $15^+$ and $16^+$ states in $^{128}$Sn. The four neutron holes in $^{128}$Sn can maximally couple to spin $16^+$ as $\nu 0h_{11/2}^{-4}$, and hence there are no $17^+$ states within the used model space. According to the shell-model calculations, the first $15^+$ and $16^+$ states in $^{128}$Sn would lie 1200 keV and 1020 keV above the $15^-$ state, which the shell model places at 4017 keV. The shell model predicts that the first $15^+$ state in $^{128}$Sn consists 97.1~\% of the $\nu 0g_{7/2}^{-1}1d_{3/2}^{-1}0h_{11/2}^{-2}$ configuration. Therefore, the beta decay from the $16^+$ isomer to the $15^+$ state would convert the $0g_{9/2}$ proton hole into $0g_{7/2}$ neutron hole, in agreement with an allowed beta decay. The calculated $16^+$ state in $^{128}$Sn is 99.5~\% $(\nu 0h_{11/2})^{-4}$ and so not likely to be fed in this beta decay. Although the observed 1280 keV transition would match in energy with the shell-model prediction for the $15^+\rightarrow 15^-$ transition, the intensity is rather low to explain the observed intensities below the $15^-$ state. We therefore assume that the 1779 keV transition feeds the $15^-$ isomer (based on its intensity and non-coincidence with other gamma transitions as reported in \cite{Fogelberg1979}) and originates from a $(15^+)$ level directly fed in the beta decay of $^{128}$In$^{m2}$. This would place the new $(15^+)$ state at 5878 keV. The right of Fig.~\ref{fig:sn128levels} shows that the shell model predicts the $15^+$ and $16^+$ states at somewhat lower energies. However, the next $15^+$ and $16^+$ states are calculated at energies greater than 7.2 MeV.

For $^{130}$In (see Fig.~\ref{fig:130inshell}), the shell model predicts a low-lying isomeric $10^-$ state but at 264 keV. This is around 200 keV higher than the experimental value of $58.6(82)$~keV. The $10^-$ isomeric state has the $(\pi 0g_{9/2}^{-1})\otimes(\nu 0h_{11/2}^{-1}$) configuration which is also the largest component for the $1^-$ ground state with $\approx 80$\% contribution. The other two significant contributions to the ground state come from the configurations $(\pi 1p_{3/2}^{-1})\otimes(\nu 1d_{5/2}^{-1}$), $\approx 8$~\%, and $(\pi 1p_{1/2}^{-1})\otimes(\nu 2s_{1/2}^{-1}$), $\approx 6$~\%. The shell-model calculations predict $3^+$ and $5^+$ states at 457 keV and 550 keV. The excitation energy for the $(5^+)$ isomer, $385.5(50)$~keV, falls below the observed $(3^+)$ state at 388.3(2)~keV \cite{Dillmann2003,Scherillo2004}. Although the $(5^+)$ state is around 200 keV lower than predicted, the experimental and theoretical spectra are in a relatively good agreement indicating that the current theoretical understanding of this mass region is reasonable. Therefore, one can expect that the theoretical predictions, such as the spin-parity of the new $^{128}$In isomer, are reliable. 

In this work isomeric states in $^{128}$In and $^{130}$In were studied with the JYFLTRAP Penning trap at the IGISOL facility. Furthermore, more accurate ground-state mass values, important for the astrophysical rapid neutron capture process and mass models, were obtained. Employing novel ion manipulation techniques, different states were resolved and masses of six beta-decaying states were measured. JYFLTRAP was also used to select the ions of interest for post-trap decay spectroscopy enabling background-free studies of the states in question. A new isomeric state in $^{128}$In feeding indirectly the $15^-$ isomer in $^{128}$Sn was discovered. Large-scale shell-model calculations suggest that this new isomeric state has a spin-parity of $16^+$ following well the systematics in the region. The shell-model calculations predict that the first isomeric state in $^{128}$In would be $10^-$, similar to $^{130}$In, but in disagreement with experiments suggesting it is $(8^-)$. The excitation energy determined here for the $(8^-)$ isomer in $^{128}$In ($285.1(25)$ keV) is somewhat lower than the theoretical prediction (452 keV). In $^{130}$In, the energy difference for the $(10^-)$ and $1^{(-)}$ states, stemming from parallel/antiparallel coupling of $(\pi 0g_{9/2}^{-1})\otimes(\nu 0h_{11/2}^{-1})$ has been found to be 58.6(82)~keV, which is around 200~keV lower than predicted by the shell model. Precise information on the energies of excited states determined in this work is crucial for producing new improved effective interactions for the nuclear shell model description of nuclei near $^{132}$Sn. Here we have demonstrated that such previously challenging isomeric states can be studied, or even new isomers discovered, using a novel combination of ion-trapping techniques and decay spectroscopy. This provides new possibilities for future studies of isomeric states. 

\section*{Acknowledgments}
This work has been supported by the EU Horizon 2020 research and innovation program under grant No. 771036 (ERC CoG MAIDEN). The support from the Academy of Finland under the Finnish Centre of Excellence Programme 2012-2017 (Nuclear and Accelerator Based Physics Research at JYFL) and projects No. 306980, 312544, 275389, 284516, 295207 is greatfully acknowledged. UK authors were supported by STFC Grant Nos. ST/L005840/1, ST/P003982/1 and ST/P005314/1.

\section*{References}

\bibliography{inbib}

\end{document}